\documentclass[superscriptaddress,twocolumn,showpacs,prb]{revtex4-1}
\usepackage[utf8]{inputenc} 
\usepackage{amsmath}
\usepackage{braket}
\usepackage{graphicx}
\usepackage{amsfonts}
\usepackage{pgfplots}
\usepackage{csquotes}
\usepackage{hhline}
\usepackage{amssymb}
\usepackage{listings}
\usepackage{color}

\makeatletter
\newcommand{\ssymbol}[1]{^{\@fnsymbol{#1}}}
\makeatother

\usepackage{lipsum}
\definecolor{codegreen}{rgb}{0,0.6,0}
\definecolor{codegray}{rgb}{0.5,0.5,0.5}
\definecolor{codepurple}{rgb}{0.58,0,0.82}
\definecolor{backcolour}{rgb}{0.95,0.95,0.92}
 
\lstdefinestyle{mystyle}{
    backgroundcolor=\color{backcolour},   
    commentstyle=\color{codegreen},
    keywordstyle=\color{magenta},
    numberstyle=\tiny\color{codegray},
    stringstyle=\color{codepurple},
    basicstyle=\footnotesize,
    breakatwhitespace=false,         
    breaklines=true,                 
    captionpos=b,                    
    keepspaces=true,                 
    numbers=left,                    
    numbersep=5pt,                  
    showspaces=false,                
    showstringspaces=false,
    showtabs=false,                  
    tabsize=2
}
 
\usepackage{subfigure}

\usepackage{dcolumn}
\usepackage{tabularx}
\usepackage[colorlinks=true,linkcolor=blue,citecolor=blue,urlcolor=blue]{hyperref}
\usepackage{longtable}
\usepackage{braket}
\usepackage{float}
\newcolumntype{C}{>{\centering\arraybackslash}X}

\begin{document}
\title{Deterministic hierarchical remote state preparation of a two-qubit entangled state using Brown \textit{et al.} state in a noisy environment}

\author{Subhashish Barik$^{a}$}
\email{subhashishbarik1995@gmail.com}
\thanks{The authors have equally contributed to this work}
\affiliation{Department of Physical Sciences,\\ Indian Institute of Science Education and Research Kolkata, Mohanpur 741246, West Bengal, India}
\author{Aakash Warke$^{a}$}
\email{aw8169@bennett.edu.in}
\thanks{The authors have equally contributed to this work}
\affiliation{Department of Physics,\\ Bennett University, Greater Noida 201310, India}

\author{Bikash K. Behera}
\email{bikash@bikashsquantum.com}
\affiliation{Bikash's Quantum (OPC) Pvt. Ltd., Balindi, Mohanpur 741246, West Bengal, India}
\affiliation{Department of Physical Sciences,\\ Indian Institute of Science Education and Research Kolkata, Mohanpur 741246, West Bengal, India}
\author{Prasanta K. Panigrahi}
\email{pprasanta@iiserkol.ac.in}
\affiliation{Department of Physical Sciences,\\ Indian Institute of Science Education and Research Kolkata, Mohanpur 741246, West Bengal, India}

\begin{abstract}
Quantum communication is one of the cutting-edge research areas today, where the scheme of Remote State Preparation (RSP) has caught significant attention of researchers. A number of different schemes of RSP have already been proposed so far. We propose here a hierarchical RSP protocol for sending a two-qubit entangled state using a seven-qubit highly entangled state derived from Brown \textit{et al.} state. We have also studied here the effects of two well known noise models namely amplitude damping (AD) and phase damping (PD) that affect the quantum communication channel used for the protocol. An investigation on the variation of fidelity of the state 
with respect to the noise operator and the receiver is made. PD noise is found to affect the fidelity more than the AD noise and the higher power receiver, obtains the state with higher fidelity than the lower
power receiver under the effect of noise. To the best of our knowledge, we believe that we have achieved the highest fidelity for the higher power receiver, 0.89 in the presence of maximum AD noise and 0.72 in the presence of maximum PD noise, compared to all the previously proposed RSP protocols in noisy environments. The study of noise is described in a very pedagogical manner for better understanding of the application of noise models to a communication protocol.
\end{abstract}

\begin{keywords}{Quantum Communication, Hierarchical Remote State Preparation, Higher power receiver, Lower power receiver}\end{keywords}

\maketitle
\section{Introduction}
Remote State Preparation (RSP) is considered to be a more efficient scheme of teleportation of a known state than the usual teleportation scheme as it involves comparatively lesser number of classical bits \cite{PPRA2000}. Ever since the introduction of Remote State Preparation, many different modified schemes of RSP \cite{BDSSTWPRL2001, XSSJPB2007, KAKJPRA2005, MCLIJTP2018, PMKMPQIP2011, PKMPJP2009, MPPRA2008, MPPRA2008(2), LWPLA2003, AHPRA2003, LSPRL2003, SPPLA2013, NKJPB2008, CNANS2014, STPQIP2017} have been reported such as hierarchical RSP \cite{SPPLA2013}, joint RSP \cite{NKJPB2008} etc. Researchers have also reported schemes such as controlled bidirectional RSP \cite{CNANS2014}, hierarchical joint remote state preparation \cite{STPQIP2017}. Since no quantum channel is practically error-free, in general every scheme works in a noisy environment. Detailed analysis of the effect of noise on a quantum channel has also been reported in the last few years \cite{STPQIP2017, SSBPQIP2015}. Motivated from these schemes, we have designed a protocol for hierarchical RSP of a two-qubit entangled state using a maximally entangled seven-qubit state obtained from Brown \textit{et al.} state. The proposed scheme is also analyzed in presence of a noisy environment.

Briefly, we can describe the protocol as follows. Alice prepares a five-qubit maximally entangled state called Brown \emph{et al.} state. Then with the addition of two ancilla qubits and by operating $CNOT$ operation on them she prepares a seven-qubit entangled state. Out of these seven entangled qubits she keeps the first one with her and sends the rest to the others; two qubits to each of Bob, Charlie and David. This entangled state can be represented now in such a fashion that to obtain the desired state, Bob would need to know the measurement outcomes of Alice and Charlie/David whereas Charlie (David) would need to know the measurement outcomes of Alice, Bob and David (Charlie). Once the receiver obtains the information about others' measurement outcomes his task is to apply certain Pauli and $CNOT$ operations to retrieve the desired state. Hence, it is easier for Bob to retrieve the information than Charlie/David. This forms the hierarchy among the two. The details of the protocol up until here is discussed in Section \ref{RSP_Sec2}. Practically, it is inevitable to do a quantum communication in a noiseless environment. Thus, we should also see how noise affects our HRSP protocol. In Section \ref{RSP_Sec3}, we make a detailed discussion on how noise affects our quantum channel through which entanglement is shared. Finally we give the explanations for the observed effects of noise and conclude in Section \ref{RSP_Sec4}. 

\section{Scheme of hierarchical remote state preparation using a seven-qubit entangled state\label{RSP_Sec2}}

Following the convention in quantum communication, we assume that the participants involved in this scheme are Alice, Bob, Charlie and David. Further, Alice wants to prepare a known state remotely at Bob's, Charlie's or David's end. We can divide our scheme to the following steps.\\

\textbf{Step 1.} \textbf{Preparation of the entangled state}\\

Alice starts with a five-qubit entangled state called \textit{Brown \emph{et al.}} state which is considered to be a maximally entangled state showing high degree of entanglement:
\begin{eqnarray}
    \ket{\psi}_{Br}&=&\frac{1}{2}(\ket{001}\ket{\phi^-}
    +\ket{010}\ket{\psi^-}\nonumber\\
    &&+\ket{100}\ket{\phi^+}+\ket{111}\ket{\psi^+})
\label{RSP_Eq1}    
\end{eqnarray}
where, $\ket{\psi^\pm}$=$\frac{1}{\sqrt{2}}(\ket{00}\pm\ket{11})$ 
and $\ket{\phi^\pm}$=$\frac{1}{\sqrt{2}}(\ket{01}\pm\ket{10})$. 

On expanding the state $\ket{\psi}_{Br}$ fully, we obtain a five-qubit cluster state of eight terms:
\begin{eqnarray}
    \ket{\psi}_{Br}&=&\frac{1}{2\sqrt{2}}(\ket{00101}+\ket{00110}+\ket{01000}\nonumber\\
    &+&\ket{01011}+\ket{10001}+\ket{10010}\nonumber\\
    &+&\ket{11100}+\ket{11111})
\label{RSP_Eq2}    
\end{eqnarray}

She then uses two ancilla qubits each prepared in state $\ket{0}$ and make them entangled with $\ket{\psi}_{Br}$ by applying two $CNOT$ operations taking the fourth and fifth qubits of $\ket{\psi}_{Br}$ as control qubits and the ancilla qubits as the target qubits:

\begin{eqnarray}
    \ket{\Psi}=CNOT (\ket{\psi}_{Br}\otimes\ket{00})
\label{RSP_Eq3}    
\end{eqnarray}

This is a very straightforward operation and after its execution Alice finally prepares the state:

\begin{eqnarray}
    \ket{\Psi}&=&\frac{1}{2\sqrt{2}}(\ket{0010101}+\ket{0011010}+\ket{0100000}\nonumber\\
    &+&\ket{0101111}+\ket{1000101}+\ket{1001010}\nonumber\\
    &+&\ket{1110000}+\ket{1111111})
\label{RSP_Eq4}    
\end{eqnarray}

We believe, this prepared state is highly entangled, if not maximally entangled, because firstly it is prepared from a maximally entangled Brown \textit{et al.} state which itself incorporates two-qubit maximally entangled Bell states and secondly it is prepared by the implementation of $CNOT$ operation which is believed to add entanglement to a state. After obtaining this state, Alice keeps the first qubit with herself and sends the second and third qubits to Bob, fourth and fifth qubits to Charlie, and sixth and seventh qubits to David through `different quantum channels'. The word `different' holds some significance here as discussed in Section \ref{RSP_Sec3}. This state is further used by Alice for the execution of HRSP of a particular two-qubit state to Bob or Charlie or David as described below.\\

\textbf{Step 2. Factorizing the state $\ket{\Psi}$ into different bases}\\

Let us assume that Alice wants to communicate the state $\ket{\xi}= \alpha\ket{00}+\beta\ket{11}$ where the relation $|\alpha|^2+|\beta|^2=1$  takes care of the normalization of the state $\ket{\xi}$. The HRSP of known state $\ket{\xi}$ mainly requires the factorization of $\ket{\Psi}$ in a particular fashion which ensures that only the sender (in this case Alice) uses a basis constructed out of the known parameters i.e., by using the information of the known state. It can be shown that $\ket{\Psi}$ can be written in the following fashion:

\begin{eqnarray}
\Ket{\Psi}_{ABCD} = \frac{1}{\sqrt{2}}\bigg(\ket{\zeta_{1}}_{A}\ket{\Xi_{1}}_{BCD}+\ket{\zeta_{2}}_{A}\ket{\Xi_{2}}_{BCD}\bigg)
\hspace{10mm}
\label{eqn5}
\end{eqnarray}

where,

\begin{eqnarray}
\ket{\Xi_{1}}&=&\frac{1}{2}\Big[\big(\alpha\Ket{01}+\beta\Ket{00}\big)_{B}\Ket{01}_{C}\Ket{01}_{D}\nonumber\\&+&\big(\beta\Ket{00}-\alpha\Ket{01}\big)_{B}\Ket{10}_{C}\Ket{10}_{D}\nonumber\\&+&\big(\alpha\Ket{10}+\beta\Ket{11}\big)_{B}\Ket{00}_{C}\Ket{00}_{D}\nonumber \\&+& \big(\beta\Ket{11}-\alpha\Ket{10}\big)_{B}\Ket{11}_{C}\Ket{11}_{D}\Big]
\label{eqn6}
\end{eqnarray}

\begin{eqnarray}
\ket{\Xi_{2}}&=&\frac{1}{2}\Big[\big(\beta\Ket{01}-\alpha\Ket{00}\big)_{B}\Ket{01}_{C}\Ket{01}_{D}\nonumber\\&-&\big(\alpha\Ket{00}+\beta\Ket{01}\big)_{B}\Ket{10}_{C}\Ket{10}_{D}\nonumber\\&+&\big(\beta\Ket{10}-\alpha\Ket{11}\big)_{B}\Ket{00}_{C}\Ket{00}_{D}\nonumber\\ &-& \big(\alpha\Ket{11}+\beta\Ket{10}\big)_{B}\Ket{11}_{C}\Ket{11}_{D}\Big]
\label{eqn7}
\end{eqnarray}
and,
\begin{eqnarray}
\ket{\zeta_{1}}=\alpha\ket{0}+\beta\ket{1}\nonumber\\
\ket{\zeta_{2}}=\beta\ket{0}-\alpha\ket{1}
\label{eqn8}
\end{eqnarray}

The subscripts $A, B, C$ and $D$ denote the qubits which belong to Alice, Bob, Charlie and David respectively. In a different fashion, $\ket\Psi$ can also be written as:

\begin{eqnarray}
  \Ket{\Psi}_{ABCD} = \frac{1}{\sqrt{2}}\bigg(\ket{\zeta_{1}}_{A}\ket{\phi_{1}}_{BCD}+\ket{\zeta_{2}}_{A}\ket{\phi_{2}}_{BCD}\bigg)
 \label{eqn9}
\end{eqnarray}

where, $\ket{\zeta_{1}}$ and $\ket{\zeta_{2}}$ are the same as given in Eq. \eqref{eqn8} and, 
\begin{eqnarray}
\ket{\phi_{1}}&=&\frac{1}{4}\Big[\Ket{++}_{B}\Ket{++}_{C}(\alpha\Ket{-+}+\beta\Ket{++})_{D}\nonumber\\ &+& \Ket{+-}_{B}\Ket{++}_{C}(\alpha\Ket{+-}-\beta\Ket{--})_{D}\nonumber\\&-&\Ket{-+}_{B}\Ket{++}_{C}(\alpha\Ket{+-}-\beta\Ket{--})_{D}\nonumber\\&+&\Ket{--}_{B}\Ket{++}_{C}(\beta\Ket{++}-\alpha\Ket{-+})_{D}\nonumber\\&+&\Ket{++}_{B}\Ket{+-}_{C}(\alpha\Ket{--}+\beta\Ket{+-})_{D}\nonumber\\&+&\Ket{++}_{B}\Ket{-+}_{C}(\alpha\Ket{++}+\beta\Ket{-+})_{D}\nonumber\\&+&\Ket{++}_{B}\Ket{--}_{C}(\alpha\Ket{+-}+\beta\Ket{--})_{D}\nonumber\\&+&\Ket{+-}_{B}\Ket{+-}_{C}(\alpha\Ket{+-}-\beta\Ket{-+})_{D}\nonumber\\&+&\Ket{+-}_{B}\Ket{-+}_{C}(\alpha\Ket{--}-\beta\Ket{+-})_{D}\nonumber\\&+&\Ket{+-}_{B}\Ket{--}_{C}(\alpha\Ket{-+}-\beta\Ket{++})_{D}\nonumber\\&-&\Ket{-+}_{B}\Ket{+-}_{C}(\alpha\Ket{++}+\beta\Ket{-+})_{D}\nonumber\\&-&\Ket{-+}_{B}\Ket{-+}_{C}(\alpha\Ket{--}+\beta\Ket{+-})_{D}\nonumber\\&-&\Ket{-+}_{B}\Ket{--}_{C}(\alpha\Ket{-+}+\beta\Ket{++})_{D}\nonumber\\&+&\Ket{--}_{B}\Ket{+-}_{C}(\beta\Ket{+-}-\alpha\Ket{--})_{D}\nonumber\\&+&\Ket{--}_{B}\Ket{-+}_{C}(\beta\Ket{-+}-\alpha\Ket{++})_{D}\nonumber\\& +&\ket{--}_{B}\ket{--}_{C}(\beta\ket{--}-\alpha\ket{+-}_{D})\Big]
\label{eqn10}
\end{eqnarray}

\begin{eqnarray}
\ket{\phi_{2}}&=&\frac{1}{4}\Big[\ket{++}_{B}\ket{++}_{C}(\beta\ket{-+}-\alpha\ket{++})_{D}\nonumber\\&+&\ket{+-}_{B}\ket{++}_{C}(\alpha\ket{--}+\beta\ket{+-})_{D}\nonumber\\&+&\ket{-+}_{B}\ket{++}_{C}(\alpha\ket{--}-\beta\ket{+-})_{D}\nonumber\\&-&\ket{--}_{B}\ket{++}_{C}(\alpha\ket{++}+\beta\ket{-+})_{D}\nonumber\\&+&\ket{++}_{B}\ket{+-}_{C}(\beta\ket{--}-\alpha\ket{+-})_{D}\nonumber\\&+&\ket{++}_{B}\ket{-+}_{C}(\beta\ket{-+}-\alpha\ket{-+})_{D}\nonumber\\&+&\ket{++}_{B}\ket{--}_{C}(\beta\ket{+-}-\alpha\ket{--})_{D}\nonumber\\&+&\ket{+-}_{B}\ket{+-}_{C}(\beta\ket{++}+\alpha\ket{-+})_{D}\nonumber\\&+&\ket{+-}_{B}\ket{-+}_{C}(\beta\ket{--}+\alpha\ket{+-})_{D}\nonumber\\&+&\ket{+-}_{B}\ket{--}_{C}(\alpha\ket{++}+\beta\ket{-+})_{D}\nonumber\\&+&\ket{-+}_{B}\ket{+-}_{C}(\alpha\ket{-+}-\beta\ket{++})_{D}\nonumber\\&+&\ket{-+}_{B}\ket{-+}_{C}(\alpha\ket{+-}-\beta\ket{--})_{D}\nonumber\\&+&\ket{-+}_{B}\ket{--}_{C}(\alpha\ket{++}-\beta\ket{-+})_{D}\nonumber\\&-&\ket{--}_{B}\ket{+-}_{C}(\alpha\ket{+-}+\beta\ket{--})_{D}\nonumber\\&-&\ket{--}_{B}\ket{-+}_{C}(\alpha\ket{-+}+\beta\ket{++})_{D}\nonumber\\&-&\ket{--}_{B}\ket{--}_{C}(\alpha\ket{--}+\beta\ket{+-})_{D}\Big]
\label{eqn11}
\end{eqnarray}

where, $A, B, C, D$, $\ket{\zeta_{1}}$ and $\ket{\zeta_{2}}$ mean the same thing as mentioned previously.\\

\textbf{Step 3. Measuring in specific basis and executing HRSP}\\

In previous subsection, we can see two different factorization of $\ket{\Psi}$ given by Eq. (\ref{eqn5}) and Eq. (\ref{eqn9}). Referring to the factorization given in Eq. (\ref{eqn5}), if Alice measures in $\{\ket{\zeta_{1}},\ket{\zeta_{2}}\}$ basis and Charlie and David measure in $\{\ket{00, \ket{01},\ket{10},\ket{11}}\}$ basis then Bob can easily retrieve the state $\ket{\xi}$ by knowing the outcomes of Alice, Charlie and David. For example, suppose Alice measures her qubit to be in state $\ket{\zeta_{1}}$ and convey her results to Bob using some classical communication. After knowing Alice's outcome, Bob figures out that his qubits along with Charlie's and David's have collapsed to the three-qubit state $\ket{\Xi_{1}}$. Now, we can see that in both state $\ket{\Xi_{1}}$ and state $\ket{\Xi_{2}}$ there is a direct correlation between Charlie's and David's measurement outcomes for all possibilities i.e., every time Charlie and Bob would have the same measurement outcomes. Suppose now, Charlie measures his qubits to be in state $\ket{01}$ and conveys the result to Bob over some classical communication channel. Knowing Charlie's outcome, Bob immediately gets to know that David would also get the same outcome as per the factorization of state $\ket{\Xi_{1}}$. This ensures Bob that he has got the state $\alpha\ket{01}+\beta\ket{00}$ and using some simple one-qubit and two-qubit operations on it he can retrieve the required state $\ket{\xi}=\alpha\ket{00}+\beta\ket{11}$. Thus, Bob needs to know the measurement outcomes of only two parties; Alice's outcome plus either Charlie's or David's outcome. The specific sets of operations that Bob has to apply on his obtained state to retrieve $\ket{\xi}$ after knowing the outcomes from Alice and Charlie/David is given in Table \ref{TableI}.

\begin{table*}
\centering
\begin{tabular}{|c|c|c|} 
\hline
Alice's outcome & Charlie/David's outcome & Bob's Operations\\
\hline
$\zeta_1$ & $\Ket{01}\Ket{01}$ & $X_{2}CX_{2-1}$\\
$\zeta_1$ & $\Ket{10}\Ket{10}$ & $iY_{2}CX_{2-1}$\\
$\zeta_1$ & $\Ket{00}\Ket{00}$ & $X_{1}CX_{2-1}$\\
$\zeta_1$ & $\Ket{11}\Ket{11}$ & $-iY_{2}X_{1,2}CX_{2-1}$\\
$\zeta_2$ & $\Ket{01}\Ket{01}$ & $iY_{2}X_{2}CX_{2-1}$\\
$\zeta_2$ & $\Ket{10}\Ket{10}$ & $-iY_{1}RCX_{2-1}$\\
$\zeta_2$ & $\Ket{00}\Ket{00}$ & $Z_{2}X_{1,2}CX_{2-1}$\\
$\zeta_2$ & $\Ket{11}\Ket{11}$ & $iY_{1}X_{2}CX_{2-1}$\\
\hline
\end{tabular}
\caption{The operations that Bob has to operate on his qubits to retrieve state $\ket{\xi}$ depending on the outcomes of Alice and Charlie/David is listed here. It can be conspicuously seen that Charlie and David get the same outcome for their measurement each time. This is the reason it suffices for David to know the outcome of only one of the two parties i.e., either Bob or Charlie in addition to the outcome of Alice. In other words, Bob would need the collaboration of only two parties (Alice plus Charlie/David). The notation $O_{1}$ (where $O$ $\in$ $\{H, X, iY, Z, CX, RCX\}$) indicates that Bob has to apply the operation $O$ on his first qubit and the notation $O_{2}$ indicates that Bob has to apply the operation $O$ on his second qubit. Similarly, $O_{1,2}$ indicates that Bob has to apply the operation $O$ on both his qubits. The operator $O_{1-2}$ indicates that Bob has to apply the operator $O$ taking the first one as the controlled qubit and the second one as the  target qubit. $CX$ and $RCX$ refers to $CNOT$ and $ReverseCNOT$ operations respectively.}
\label{TableI}
\end{table*}

Let us now refer to the factorization given in Eq. (\ref{eqn9}). We can see that here, the state is finally retrieved by David. Suppose, Alice measures in the $\{\ket{\zeta_{1}},\ket{\zeta_{2}}\}$ basis and finds her qubit to be in state $\ket{\zeta_{1}}$. Once David obtains Alice's measurement outcome over a classical channel he gets to know that his qubits along with Charlie's and David's have collapsed to the six-qubit state $\ket{\phi_{1}}$. However, we don't see any kind of correlation between the measurement outcomes of Bob and Charlie for states $\ket{\phi_{1}}$ and $\ket{\phi_{2}}$ (as we saw between Charlie and David for the state $\ket{\Xi_{1}}$ and $\ket{\Xi_{2}}$). That means to figure out which state he has obtained, David needs the measurement outcomes of both Bob and Charlie. Suppose, Bob and Charlie individually measures in the two-qubit Hadamard basis and find their states to be $\ket{+-}$ and $\ket{++}$ respectively. Once David receives this information over a classical channel from Bob and Charlie he gets to know that his state has collapsed to state $\alpha\ket{+-}-\beta\ket{--}$ as it can be seen from Eq. (\ref{eqn10}) and accordingly he will apply some simple one-qubit and two-qubit operations on his qubits to obtain the desired state $\ket{\xi}$= $\alpha\ket{00}+\beta\ket{11}$. The exact set of operations David has to apply based on the outcomes of Alice, Bob and Charlie is given in Table \ref{table2} and Table \ref{table3}. In a similar fashion, we can also factorize $\ket{\phi_{1}}$ and $\ket{\phi_{2}}$ in such a way that Charlie receives the desired state for which he will have to know the outcomes of Alice, Bob and David. 

\begin{table*}
\begin{tabular}{|c|c|c|} 
\hline
Alice's outcome & Bob/Charlie's  outcome & David's Operations \\
\hline
$\zeta_{1}$ & $\Ket{++}_{B}\Ket{++}_{C}$ & $H_{1,2}X_{1}CX_{1-2}$ \\
$\zeta_{1}$ & $\Ket{+-}_{B}\Ket{++}_{C}$ & $H_{1,2}Z_{1}RCX_{1-2}$ \\
$\zeta_{1}$ & $\Ket{-+}_{B}\Ket{++}_{C}$ & $H_{1,2}Z_{2}RCX_{1-2}$ \\
$\zeta_{1}$ & $\Ket{--}_{B}\Ket{++}_{C}$ & $H_{1,2}iY_{1}CX_{1-2}$ \\
$\zeta_{1}$ & $\Ket{++}_{B}\Ket{+-}_{C}$ & $H_{1,2}X_{1,2}CX_{1-2}$\\
$\zeta_{1}$ & $\Ket{++}_{B}\Ket{-+}_{C}$ & $H_{1,2}CX_{1-2}$\\
$\zeta_{1}$ & $\Ket{++}_{B}\Ket{--}_{C}$ & $H_{1,2}RCX_{1-2}$\\
$\zeta_{1}$ & $\Ket{+-}_{B}\Ket{+-}_{C}$ & $H_{1,2}Z_{1}CX_{1-2}$\\
$\zeta_{1}$ & $\Ket{+-}_{B}\Ket{-+}_{C}$ & $H_{1,2}X_{1,2}Z_{1}CX_{1-2}$\\
$\zeta_{1}$ & $\Ket{+-}_{B}\Ket{--}_{C}$ & $H_{1,2}X_{1}Z_{1}CX_{1-2}$ \\
$\zeta_{1}$ & $\Ket{-+}_{B}\Ket{+-}_{C}$ & $H_{1,2}X_{2}Z_{2}RCX_{1-2}$ \\
$\zeta_{1}$ & $\Ket{-+}_{B}\Ket{-+}_{C}$ & $H_{1,2}iY_{2}X_{1}CX_{1-2}$ \\
$\zeta_{1}$ & $\Ket{-+}_{B}\Ket{--}_{C}$ & $H_{1,2}X_{2}Z_{2}X_{1}X_{2}CX_{1-2}$ \\ 
$\zeta_{1}$ & $\Ket{--}_{B}\Ket{+-}_{C}$ & $H_{1,2}iY_{1}X_{2}CX_{1-2}$ \\ 
$\zeta_{1}$ & $\Ket{--}_{B}\Ket{-+}_{C}$ & $H_{1,2}X_{2}Z_{2}RCX_{1-2}$ \\ 
$\zeta_{1}$ & $\Ket{--}_{B}\Ket{--}_{C}$ & $H_{1,2}Z_{2}Z_{1}RCX_{1-2}$\\
\hline
\end{tabular}
\caption{The operations that David has to operate on his qubits to retrieve state $\ket{\xi}$ when Alice gets the outcome $\ket{\zeta_{1}}$ are listed here. It can be clearly seen that Bob and Charlie get random outcomes for their measurements with respect to each other. This is the reason David would require the collaboration of all other parties unlike that in Table \ref{TableI} to retrieve state $\ket{\xi}$. The notations $O_{1}$, $O_{2}$, $O_{1,2}$, $O_{1-2}$, $CX$ and $RCX$ have the same meaning as described in Table \ref{TableI} with the only difference that here David would apply those operations instead of Bob. }
\label{table2}
\end{table*}

\begin{table*}
\begin{tabular}{|c|c|c|} 
\hline
Alice's outcome & Bob/Charlie's outcome & David's Operations \\
\hline
$\zeta_{2}$ & $\Ket{++}_{B}\Ket{++}_{C}$ & $H_{1,2}X_{1}Z_{1}X_{1}CX_{1-2}$\\
$\zeta_{2}$ & $\Ket{+-}_{B}\Ket{++}_{C}$ & $H_{1,2}X_{1,2}CX_{1-2}$\\
$\zeta_{2}$ & $\Ket{-+}_{B}\Ket{++}_{C}$ & $H_{1,2}X_{1,2}Z_{1}CX_{1-2}$\\
$\zeta_{2}$ & $\Ket{--}_{B}\Ket{++}_{C}$ & $H_{1,2}X_{2}Z_{2}RCX_{1-2}$\\
$\zeta_{2}$ & $\Ket{++}_{B}\Ket{+-}_{C}$ & $H_{1,2}iY_{1}X_{1,2}CX_{1-2}$\\
$\zeta_{2}$ & $\Ket{++}_{B}\Ket{-+}_{C}$ & $H_{1,2}Z_{1}X_{2}CX_{1-2}$\\
$\zeta_{2}$ & $\Ket{++}_{B}\Ket{--}_{C}$ & $H_{1,2}Z_{1}X_{1,2}CX_{1-2}$\\
$\zeta_{2}$ & $\Ket{+-}_{B}\Ket{+-}_{C}$ & $H_{1,2}X_{1}CX_{1-2}$\\
$\zeta_{2}$ & $\Ket{+-}_{B}\Ket{-+}_{C}$ & $H_{1,2}X_{2}CX_{1-2}$\\
$\zeta_{2}$ & $\Ket{+-}_{B}\Ket{--}_{C}$ & $H_{1,2}CX_{1-2}$\\
$\zeta_{2}$ & $\Ket{-+}_{B}\Ket{+-}_{C}$ & $H_{1,2}iY_{1}CX_{1-2}$\\
$\zeta_{2}$ & $\Ket{-+}_{B}\Ket{-+}_{C}$ & $H_{1,2}X_{2}Z_{1}CX_{1-2}$\\
$\zeta_{2}$ & $\Ket{-+}_{B}\Ket{--}_{C}$ & $H_{1,2}Z_{1}CX_{1-2}$\\
$\zeta_{2}$ & $\Ket{--}_{B}\Ket{+-}_{C}$ & $H_{1,2}Z_{1}X_{1}CX_{2-1}$\\
$\zeta_{2}$ & $\Ket{--}_{B}\Ket{-+}_{C}$ & $H_{1,2}X_{1}CX_{1-2}$\\
$\zeta_{2}$ & $\Ket{--}_{B}\Ket{--}_{C}$ & $H_{1,2}Z_{2}CX_{1-2}$\\
\hline
\end{tabular}
\caption{The operations that David has to operate on his qubits to retrieve state $\ket{\xi}$ when Alice gets the outcome $\ket{\zeta_{2}}$ are listed here. Similar to our observation in Table \ref{table2}, Bob and Charlie get random outcomes for their measurements with respect to each other and consequently David would require the collaboration of all other parties to retrieve state $\ket{\xi}$. The notations $O_{1}$, $O_{2}$, $O_{1,2}$, $O_{1-2}$, $CX$ and $RCX$ have the same meaning as described in Table \ref{table2}.}
\label{table3}
\end{table*}

Thus we see here that, to retrieve the required state $\ket{\xi}$ Bob needs the information of only two people; (Alice and Charlie/David) where as Charlie (David) needs the information of three people; (Alice, Bob, David (Charlie). Thus it is easier for Alice to prepare a known state remotely at Bob's end than that in Charlie/David's end. In other words, Bob here is the higher power receiver where as Charlie and David are the lower power receivers having the same power. This forms a hierarchy between Bob, Charlie and David, where Bob lies at the top level followed by Charlie and David lying in the same level. Hence, this scheme becomes an example of HRSP.

\section{Effect of noise on the quantum channel \label{RSP_Sec3}}

Practically, it is impossible to do quantum communication through a noiseless channel. At best what we can do is to study the sources and effects of noise and try to minimize them. We can conveniently study the effects of different kinds of noises for the HRSP of state $\ket{\xi}$ by studying the evolution of $\rho=\ket{\Psi}\bra{\Psi}$ under the effects of different noise operators (more commonly called Kraus operators). For the sake of simplicity, we study here the effects of only two prominent kinds of noises namely amplitude damping (AD) and phase damping (PD). Fortunately, there exist the defined forms of noise operators for AD and PD noise for which we would not have to separately construct them here \cite{DMW2007,NCCUP2000}. The noise operators usually evolve $\rho$ in the following way:

\begin{eqnarray}
\rho\xrightarrow{}\sum\limits_{i} K_{i} \rho K_{i}\ssymbol{2}
\label{eqn12}
\end{eqnarray}

where $K_{i}$ is any particular noise operator. The noise operators are constructed following the operator-sum representation and the set of noise operators, \{$K_{0},K_{1}$\} for AD can be given as \cite{STPQIP2017,SSBPQIP2015}: 

\begin{eqnarray}
  K_0=
  \begin{pmatrix}
    1 & 0 \\
    0 & \sqrt{1-\eta_{A}}
  \end{pmatrix}\quad
  K_1=
  \begin{pmatrix}
    0 & \sqrt{\eta_{A}} \\
    0 & 0
  \end{pmatrix}
\label{eqn13}
\end{eqnarray}

where, $\eta_{A}$ describes the probability of error due to AD noise. Similarly, the set of noise operators, \{$E_{0},E_{1},E_{2}$\} for PD can be given as \cite{STPQIP2017,SSBPQIP2015}:

\begin{eqnarray}
  E_0=
  \begin{pmatrix}
    \sqrt{1-\eta_{P}} & 0 \\
    0 & \sqrt{1-\eta_{P}}
  \end{pmatrix}\nonumber\\
  E_1=
  \begin{pmatrix}
    \sqrt{\eta_{P}} & 0 \\
    0 & 0
  \end{pmatrix}\nonumber\\
  E_2=
  \begin{pmatrix}
      0 & 0 \\
      0 & \sqrt{\eta_{P}}
  \end{pmatrix}
\label{eqn14}
\end{eqnarray}

where, $\eta_{P}$ describes the probability of error due to PD noise. Further, $\eta_{A}$ and $\eta_{P}$ can take values between 0 to 1. Next, we see the effects of AD noise and PD noise one by one. 


\subsection{Effect of AD noise}
Referring to Eq. (\ref{eqn12}) we can describe the evolution of $\rho$ under the effect of AD noise as follows:

\begin{widetext}
\begin{equation}
    \rho' = \sum\limits_{i,j,l} \big({I^{A}_{2}}\otimes{K^{B_{0}}_{i}}\otimes{K^{B_{1}}_{i}}\otimes{K^{C_{0}}_{j}}\otimes{K^{C_{1}}_{j}}\otimes{K^{D_{0}}_{l}}\otimes{K^{D_{1}}_{l}}\big)\ \rho\ \big({I^{A}_{2}}\otimes{K^{B_{0}}_{i}}\otimes{K^{B_{1}}_{i}}\otimes{K^{C_{0}}_{j}}\otimes{K^{C_{1}}_{j}}\otimes{K^{D_{0}}_{l}}\otimes{K^{D_{1}}_{l}}\big)\ssymbol{2}
\label{eqn15}
\end{equation}
\end{widetext}

Clearly, $\rho$ is a 128$\times$128 (or $2^7\times2^7$) matrix as $\ket{\Psi}$ defines a seven-qubit system. Accordingly, we have seven one-qubit operators on the left and seven on the right of $\rho$. The superscripts on the noise operators denote the particular qubit on which the operator would operate. Furthermore, we should be clear that $B_0$ and $B_1$ denote the two qubits of Bob, $C_0$ and $C_1$ denote the two qubits of Charlie and $D_0$ and $D_1$ denote the two qubits of David. The subscript however, denotes the two different AD operators i.e., i, j, l $\in$ $\{0,1\}$. It is assumed that Alice prepares the entangled state and then distributes the qubits to Bob, Charlie and David through some quantum channels. Thus, we can practically assume that the noise would affect all qubits except for Alice's qubit. Further, assuming that the two qubits belonging to the same receiver are communicated through the same quantum channel we believe that they are affected by the same noise operators and hence we put the same index on the subscript of noise operators that acts on the qubits of the same receiver. For example, the two noise operators $K^{B_{0}}$ and $K^{B_{1}}$ acting on the two qubits of Bob have the same subscript $``i"$. Following similar argument, the noise operators acting on the qubits of two different receivers bear different index on their subscripts.

We then compute $\rho'$ using MATLAB and obtain a 128$\times$128 matrix which turns out to be a function of $\eta_{A}$. Certainly, now $\rho'$ is the state which evolves as per the measurements made by Alice, Bob and Charlie as depicted by the protocol in Section \ref{RSP_Sec2} to some state $\rho''$ (say). We construct this unitary evolution operator by carefully choosing the measurement operators that would act on each qubit. Hence,

\begin{eqnarray}
\rho'' = U \rho' U^{\dagger}
\label{eqn16}
\end{eqnarray}

where, $U$ is the following operator:
\begin{eqnarray}
U = {M_{A}}\otimes{M_{B_{0}}}\otimes{M_{B_{1}}}\otimes{M_{C_{0}}}\otimes{M_{C_{1}}}\otimes{M_{D_{0}}}\otimes{M_{D_{1}}}\nonumber\\
\label{eqn17}
\end{eqnarray}

where, $M_{X}$ represents the measurement operator that would act on the qubit X. After obtaining $\rho^{''}$, we have to normalize it, just the way states are normalized after measurement. Thus we obtain our normalized $\rho^{''}$ (say, $\rho^{'''}$) by dividing it by the trace of $\rho^{''}$:

\begin{eqnarray}
\rho''' = \frac{\rho''}{Tr(\rho'')}=\frac{U \rho' U^{\dagger}}{Tr(U \rho' U^{\dagger})}
\label{eqn18}
\end{eqnarray}

$\rho'''$ itself is a $128\times128$ matrix and in order to know what state the receiver gets we must reduce it to a $4\times4$ matrix (since the receiver obtains a two-qubit state here). If Bob is the receiver then we have to trace out the qubits of Alice, Charlie and David (corresponding to $A, C_{0}, C_{1}, D_{0}, D_{1}$):

\begin{eqnarray}
\rho^{Bob} = Tr_{A, C_{0}, C_{1}, D_{0}, D_{1}}(\rho''')
\label{eqn19}
\end{eqnarray}

Once the receiver receives this state he uses the appropriate unitary operator, say \textit{O}, to obtain the desired state $\ket{\xi}$:
\begin{eqnarray}
\rho_{n}=O\rho^{Bob}O^{\dagger}
\label{eqn20}
\end{eqnarray}

Hence, $\rho_{n}$ is the state received through the noisy channel. As noise here is parameterized by $\eta_{A}$ we can see $\rho_{n}$ to be a function of $\eta_{A}$. Now, to see the effect of noise on our desired state $\rho_{0}$=$\ket{\xi}\bra{\xi}$ we calculate the fidelity between $\rho_{0}$ and $\rho_{n}$ since fidelity would show how close is $\rho_{n}$ to $\rho_{0}$:

\begin{eqnarray}
F=Tr\bigg(\sqrt{\sqrt{\rho_{0}}\rho_{n}\sqrt{\rho_{0}}}\bigg)
\label{eqn21}
\end{eqnarray}

Let us explicitly calculate the fidelity for a particular set of measurements. Let us consider the the first set of measurements (the first line in the table) from Table \ref{TableI} where, Bob is the receiver of the desired state. It should be noted that for AD, $\rho'$ remains the same for all the sets of measurements given in Table \ref{TableI}-\ref{table3}. However, the operators $U$ and $O$ differ from set to set. The measurement operators of Alice, Charlie, David and Bob to be used in Eq. \eqref{eqn10} can be given as:

\begin{eqnarray}
M_{A}=\bigg(\frac{\ket{0}+\ket{1}}{\sqrt{2}}\bigg)\bigg(\frac{\bra{0}+\bra{1}}{\sqrt{2}}\bigg)=\frac{1}{2} \begin{pmatrix}
    1 & 1 \\
    1 & 1 \\
  \end{pmatrix}\nonumber\\
M_{C}=M_{C_{0}}\otimes M_{C_{1}}=\ket{01}\bra{01}=\begin{pmatrix}
    0 & 0 & 0 & 0 \\
    0 & 1 & 0 & 0 \\
    0 & 0 & 0 & 0 \\
    0 & 0 & 0 & 0
     \end{pmatrix}\nonumber\\
M_{D}=M_{D_{0}}\otimes M_{D_{1}}=\ket{01}\bra{01}=\begin{pmatrix}
    0 & 0 & 0 & 0 \\
    0 & 1 & 0 & 0 \\
    0 & 0 & 0 & 0 \\
    0 & 0 & 0 & 0
  \end{pmatrix}\nonumber\\
 M_{B}=M_{B_{0}}\otimes M_{B_{1}}=I_2\times I_2=\begin{pmatrix}
    1 & 0 & 0 & 0 \\
    0 & 1 & 0 & 0 \\
    0 & 0 & 1 & 0 \\
    0 & 0 & 0 & 1
  \end{pmatrix} 
\label{eqn22}
\end{eqnarray}

To simplify the calculations, we have taken $\alpha$=$\beta$=$1/\sqrt{2}$ in $M_{A}$=($\alpha\ket{0}$+$\beta\ket{1}$)($\alpha^*\bra{0}$+$\beta^*\bra{1}$) in Eq. \eqref{eqn15}. Further, we can see that Bob's measurement operator is an Identity matrix. This is because, while the state is evolving under the noise Bob performs no measurement. More precisely, Bob performs a local measurement only after he gets to know the outcomes of others i.e., only after tracing out the qubits of Alice, Charlie and David. Consequently, we take Bob's measurement into consideration only while calculating $\rho_{n}$. For the particular set of measurements that we have considered, Bob's measurement operator ($O$) as used in Eq. \eqref{eqn23} becomes:

\begin{eqnarray}
O=M_{B_{0}}\otimes M_{B_{1}}=CX_{2\rightarrow{1}}(I\otimes X) =
\begin{pmatrix}
    0 & 1 & 0 & 0 \\
    0 & 0 & 1 & 0 \\
    0 & 0 & 0 & 1 \\
    1 & 0 & 0 & 0
\end{pmatrix}\nonumber\\
\label{eqn23}
\end{eqnarray}

Using the operators given in Eq. \eqref{eqn22} we first calculate $U$ as per Eq. \eqref{eqn17} and then $\rho'''$ as per Eq. \eqref{eqn18}. Then we trace out the qubits corresponding to $A, C_{0}, C_{1}, D_{0}, D_{1}$ using a MATLAB program to find out $\rho^{Bob}$ which is a 4$\times$4 matrix. After that using the operator $O$ (a 4$\times$4 matrix as we would expect it to be multiplied with $\rho^{Bob}$) we calculate $\rho_{n}$ as per the formula in Eq. \eqref{eqn20}. We have taken, $\alpha=\beta=1/\sqrt{2}$, thus, the intended state that Bob should obtain in the absence of noise is:

\begin{eqnarray}
\rho_{0}= \frac{1}{\sqrt{2}}\bigg(\ket{00}+\ket{11}\bigg)\frac{1}{\sqrt{2}}\bigg(\bra{00}+\bra{11}\bigg)\nonumber\\
\rho_{0}=\frac{1}{2}
\begin{pmatrix}
    1 & 0 & 0 & 1 \\
    0 & 0 & 0 & 0 \\
    0 & 0 & 0 & 0 \\
    1 & 0 & 0 & 1
  \end{pmatrix}
\label{eqn24}
\end{eqnarray}

\begin{figure*}[]
    \centering
    \includegraphics[scale=0.187]{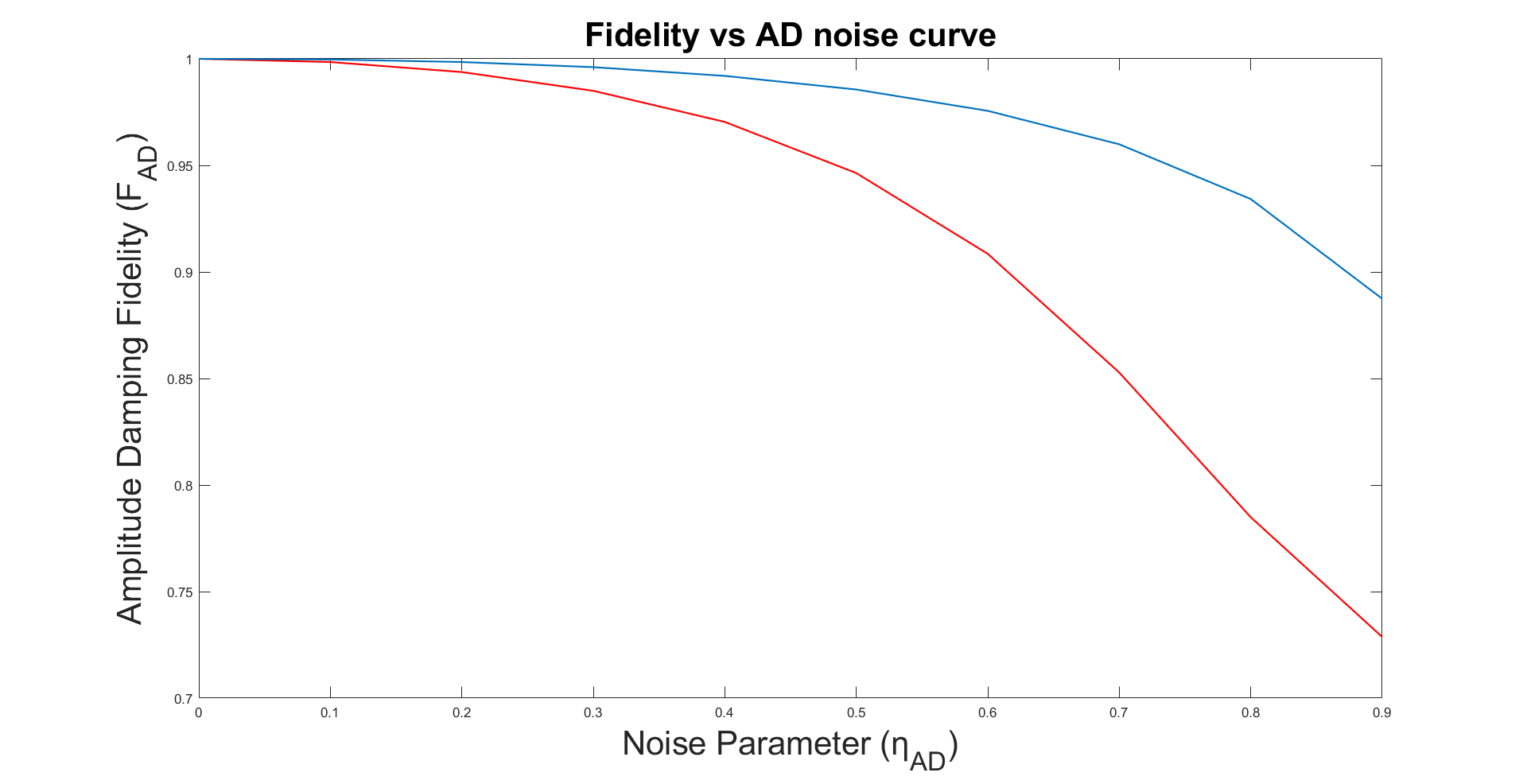}
    \caption{This figure depicts the variation of fidelity with the AD noise parameter. The blue line refers to the fidelity when the higher power receiver (Bob here) receives the state $\ket{\xi}$ and the red line refers to the fidelity when the lower power receiver (David here) receives the state $\ket{\xi}$. The fidelity curve for Bob lies above that of David. Further, the lowest fidelity for Bob is 0.89 where as that for David is 0.73. This ensures that the fidelity of the higher power receiver is more affected by the AD noise than the fidelity of the lower power receiver.}
    \label{Fig 1}
\end{figure*}

Then, finally we calculate the fidelity by plugging in $\rho$ and $\rho_{n}$ in Eq. \eqref{eqn21}. We obtain fidelity as a function of $\eta_{A}$. Further, as $\eta_{A}$ can take values between 0 and 1 only we calculate and plot fidelity as a function of $\eta_{A}$ by varying $\eta_{A}$ in steps of 0.1. This plot is shown in Fig. \ref{Fig 1}. To see the effect of noise based on the power of the receiver we calculate the fidelity once taking Bob (higher power receiver) as the receiver using the first set of measurements given in the top line of the Table \ref{TableI}, and once taking David (lower power receiver) as the receiver  using the first set of measurements given in the top line of the Table \ref{table2}. The behaviour of Bob's and David's fidelities with $\eta_{AD}$ is depicted by the blue line and red line in Fig. \ref{Fig 1} respectively. It can be understood that the behaviour of fidelities with  $\eta_{AD}$ would be the same for other sets of measurements of the particular table as long as we calculate it for the same receiver.\\ 

\subsection{Effect of PD noise}

In studying the effects of PD noise, we take the same approach as we took for AD noise. At first, we write down the evolution of $\rho$ in terms of the PD noise operators:

\begin{widetext}
\begin{equation}
    \rho' = \sum\limits_{i,j,l} \big({I^{A}_{2}}\otimes{E^{B_{0}}_{i}}\otimes{E^{B_{1}}_{i}}\otimes{E^{C_{0}}_{j}}\otimes{E^{C_{1}}_{j}}\otimes{E^{D_{0}}_{l}}\otimes{E^{D_{1}}_{l}}\big)\ \rho\ \big({I^{A}_{2}}\otimes{E^{B_{0}}_{i}}\otimes{E^{B_{1}}_{i}}\otimes{E^{C_{0}}_{j}}\otimes{E^{C_{1}}_{j}}\otimes{E^{D_{0}}_{l}}\otimes{E^{D_{1}}_{l}}\big)\ssymbol{2}
\end{equation}
\label{eqn25}
\end{widetext}

\begin{figure}[]
    \centering
    \includegraphics[scale=0.187]{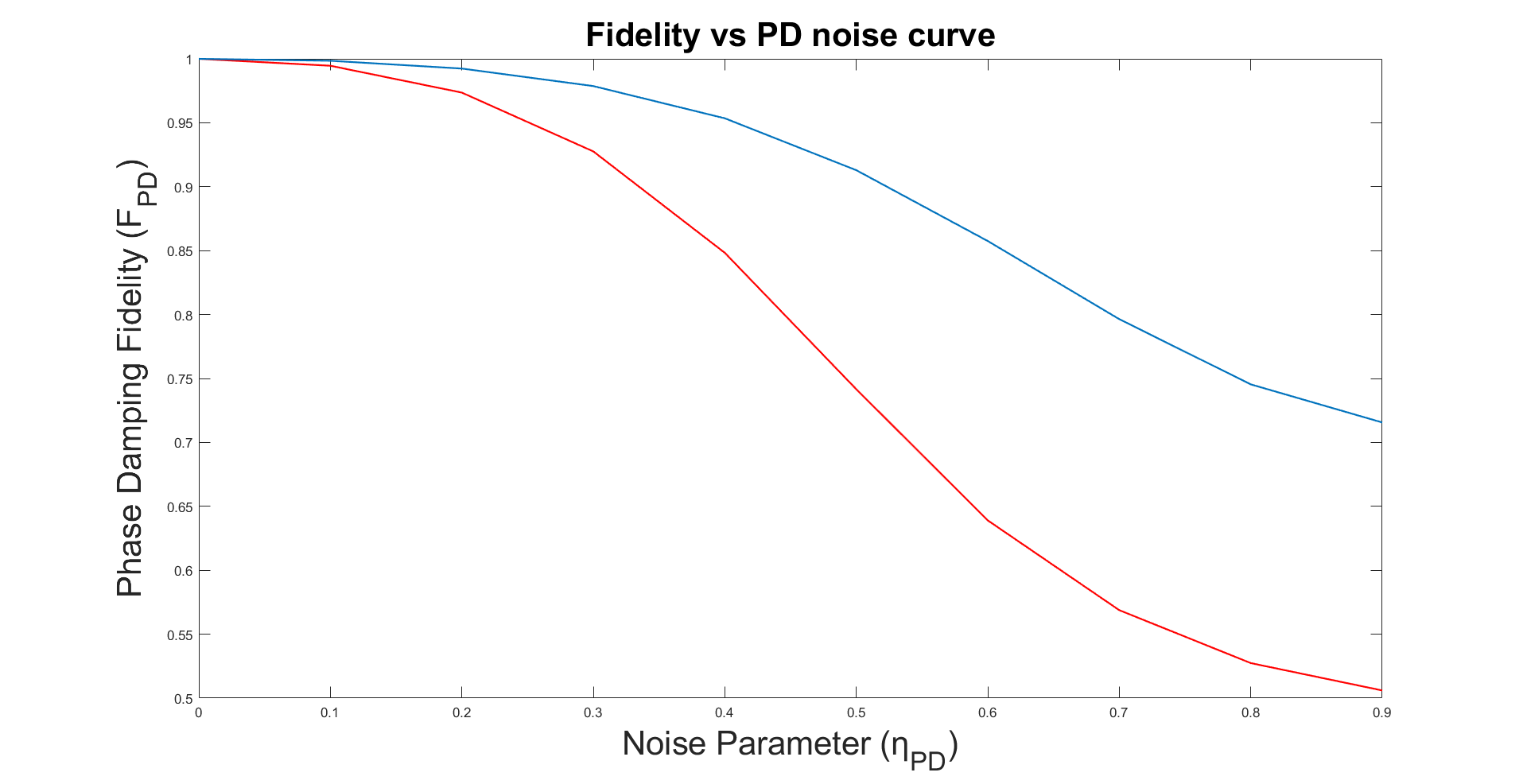}
    \caption{This figure depicts the variation of fidelity with the PD noise parameter. The blue line refers to the fidelity when the higher power receiver (Bob here) receives the state $\ket{\xi}$ and the red line refers to the fidelity when the lower power receiver (David here) receives the state $\ket{\xi}$. Just like in Fig. \ref{Fig 1} the fidelity curve for Bob lies above that of David. Further, the lowest fidelity for Bob is 0.72 where as for David is 0.50. This implies that the fidelity of the higher power receiver is more affected by the PD noise than the fidelity of the lower power receiver. Moreover, it can be noted that the fidelity is more affected by PD noise than AD noise.}
    \label{Fig 2}
\end{figure}

We calculate this $128\times 128$ matrix in MATLAB plugging in the values of the PD noise operators from equation \eqref{eqn14}. From here onward, the rest of the formalism is exactly the same with what we did for AD noise. We calculate $U$ using the sets of measurements used by Alice, Bob, Charlie and David followed by the calculation of $\rho^{Bob}$ and then $\rho_n$. Finally, we calculate the fidelity using the formula given in Eq. \eqref{eqn21}. For the first set of measurements in Table \ref{TableI}, which was also used for the AD noise, we calculate the fidelity. The fidelity comes out to be a function $\eta_{PD}$ just as we would have expected. We then plot this fidelity for different values of $\eta_{PD}$ varying between 0 and 1 as shown in Fig. \ref{Fig 2}. We calculate the fidelity once taking Bob (higher power receiver) as the receiver using the first set of measurements given in the top line of the Table \ref{TableI}, and once taking David (lower power receiver) as the receiver  using the first set of measurements given in the top line of the Table \ref{table2}. The behaviour of Bob's and David's fidelities with $\eta_{PD}$ is depicted by the blue line and red line in Fig. \ref{Fig 2} respectively. The behaviour of fidelities with $\eta_{PD}$ would be the same for the other sets of measurements of the particular table as long as we calculate it for the same receiver.

\section{Discussion \label{RSP_Sec4}}
We have proposed here a scheme for the HRSP of a two-qubit entangled state using a seven-qubit entangled state generated from Brown \textit{et al.} state. Emphasis has been put on describing a generic HRSP protocol in the most pedagogic way for better understanding of the protocol. We believe the way we have presented the protocol here would help readers to come up with the HRSP of even more complicated states (involving maximum number of qubits, say \textit{n}) using the factorization of a multipartite-entangled state involving maximum number of qubits, say \textit{m}. The efficiency of the protocol would depend upon the ratio $\frac{n}{m}$; higher being more efficient.

We have also shown the effects of AD and PD noises on our protocol considering the noise to only affect the travelling qubits (the ones Alice sends to others). From the two figures it can be seen that the least value of fidelity for AD noise is 0.73 and for PD noise is 0.5 which is quite better compared to many of the recently proposed protocols \cite{STPQIP2017, SSBPQIP2015, CLHIJP2019, ZMGIJTP2019, QXJPLA2019, FSNDIJQI2019}. This proves our scheme to be quite trustable for practical implementation. We believe, the reason behind the higher fidelity is the highly entangled seven-qubit state that we have taken. It would be interesting to know the exact amount of entanglement present in this seven-qubit state which is generated from the maximally entangled Brown \textit{et al.} state. We would like to leave it to the future researchers to measure the amount of entanglement (in any plausible measure of entanglement) in this seven-qubit entangled state. We have seen that the PD noise has more effect on the fidelity than the AD noise.

Another important conclusion from our study of noise is that the higher power agent (Bob here) would receive a high fidelity state than the low power agent (David here) as it can be very conspicuously seen in Figs. \ref{Fig 1} and \ref{Fig 2}. However, in order to generalize this observation to all different kinds of noises and all different kinds of hierarchical protocols, we need a much detailed study of different kinds of hierarchical protocols in different noisy environments. Thus, we feel that further research in this area may lead to more concrete results which would help to design more efficient protocols for practical implementations in future.

 \section{Acknowledgments}
\label{qlock_acknowledgments}
S.B. and A.W. would like to thank Bikash's Quantum (OPC) Pvt. Ltd. and IISER Kolkata for providing hospitality during the course of the project work. A.W would like to thank Bennett University for providing the license required for operating MATLAB. A.W. would also like to thank Yash Palan from IISER Bhopal for his suggestions during the calculations with MATLAB. B.K.B. acknowledges the support of Institute fellowship provided by IISER Kolkata. The authors would also like to acknowledge QuEST initiative to DST/ICPS/QUST/THEME-1/2019 for the grant support.

\end{document}